\documentstyle[12pt,epsfig]{article}
\def\beqra{\begin{eqnarray}} 
\def\eeqra{\end{eqnarray}}
\def\beq{\begin{equation}}      
\def\eeq{\end{equation}}
\def\ds{\displaystyle}

\def\L{\Lambda}
\def\re#1{(\ref{#1})}
\def\D{\Delta}
\def\G{\Gamma}
\def\p{\partial}
\def\de{\delta}
\renewcommand{\Re}{\mathop{\mathrm{Re}}}
\renewcommand{\Im}{\mathop{\mathrm{Im}}}
\newcommand{\tr}{\mathop{\mathrm{tr}}}
\newcommand{\Tr}{\mathop{\mathrm{Tr}}}

\def\i{i}

\def\d{d}
\def\e{e}
\def\half{\mbox{\small $\frac{1}{2}$}}

\begin{document}


%
%
\begin{titlepage}
\begin{flushright}
CERN-TH/96-23\\
SHEP 96-05\\
\end{flushright}
\vspace{2cm}
\centerline{\Large\bf  Wilson Renormalization Group formulation}
\vspace{10pt}
\centerline{\Large\bf  of Real Time thermal field theories}
\vspace{24pt}
\centerline{\large M. D'Attanasio}
\centerline{\it Department of Physics, University of 
Southampton, United Kingdom}
\centerline{\it
and INFN Gruppo collegato di Parma, Italy}
\vspace{20 pt}
\centerline{\large M. Pietroni}
\centerline{\it Theory Division, CERN}
\centerline{\it CH-1211 Geneva 23, Switzerland}

\vspace{24pt}
\begin{abstract}
We apply Renormalization Group techniques to the Real Time formulation 
of thermal field theory. Due to the separation between the $T=0$ and
the $T\neq 0$ parts of the propagator in this formalism, one can derive exact
evolution equations for the Green functions describing the effect of 
integrating out thermal
fluctuations of increasing wavelengths, the initial conditions being the 
renormalized Green functions of the $T=0$ theory.
As a first application, we study the phase transition for the real scalar 
theory,
computing the order of the transition, the critical temperature, and critical
exponents, in different approximations to the evolution equations for the 
scalar potential. 
\end{abstract}
\vspace{6.3cm}
CERN-TH/96-23\\
February 1996
\end{titlepage}

\section{Introduction}

The dynamics of a second order, or weakly first order, phase 
transition is governed by long wavelength fluctuations of the 
order parameter.
These fluctuations, whose length scale is much larger than the 
inverse temperature of the system, are essentially classical, 
since the probability of a quantum fluctuation over such scales 
is highly suppressed. 
For this reason, the details of the microscopic theory are not 
relevant to the description of the critical behaviour of the system, 
which can be successfully studied by using classical models, 
such as the Ising model for ferromagnets or the Ginsburg-Landau theory 
for superconductors.

In these models, a free energy can be defined as a functional of a
macroscopic order parameter, which is allowed to vary only on large 
length scales.
The effect of the short wavelength (quantum and thermal) fluctuations is
incorporated by the parameters appearing in the free energy (masses, coupling
constants, etc.), which are usually treated as phenomenological parameters but
should in principle be computable, starting from the underlying theory. 

The Wilson Renormalization Group (RG) \cite{Wilson} provides the natural
framework in which this procedure can be systematically carried out. The main 
idea is to start from a microscopic theory, the parameters of which are
supposed to be known, and then progressively integrate out the high frequency
modes of the order parameter down to some infra-red cutoff $\Lambda$.  
In this way one obtains a coarse-grained order parameter and the corresponding 
effective action, which can be used as the relevant tool to describe 
the system. 

In this paper, we will apply this idea to quantum field theory at finite 
temperature.
Our approach will be the following. First, we will assume that we 
``know'' the zero-temperature renormalized quantum field theory, 
which means that there exists
some
reliable approximation method (perturbation theory, lattice simulations, 
etc.) to
compute the zero-temperature renormalized Green functions. In this way, the
renormalization constants can be fixed by the experimental measurements and
all the parameters of the theory are known. In a RG language, we assume that
all the quantum fluctuations have already been  integrated out.

Second, we will
integrate out thermal modes only, for frequencies higher than the infrared
cutoff $\Lambda$. For non-vanishing values for $\Lambda$, we will obtain
coarse-grained order parameter and free energy, which are the appropriate
objects to study the dynamics of long wavelength thermal fluctuations. 
In the limit  $\Lambda \rightarrow 0$, we will obtain the finite
temperature quantum field theory in thermal equilibrium.

In this approach, since all the quantum fluctuations are integrated out from the
beginning, it will be possible to relate the thermal field theory and the
physical (renormalized) quantum field theory at zero temperature in a 
transparent way.

The ideal framework to perform this program of coarse-graining of 
thermal fluctuations is the Real Time (RT) formulation of thermal 
field theories \cite{Niemi}, in which the thermal part in the free 
propagators is well separated from the zero-temperature quantistic one. 
As is well known, the price to pay for this is a doubling of the number 
of degrees of freedom.
These ``ghost'' fields are necessary to cancel the so-called pinch 
singularities, which are due to the fact that the thermal fluctuations 
are on-shell. 
Because of these technical complications Matsubara's Imaginary Time 
(IT) approach
\cite{Matsubara} is more popular in the literature. 

In particular, the coarse-graining procedure can be precisely 
formulated in the so-called Closed-Time-Path (CTP) formalism 
\cite{noneq} (see refs.~\cite{Calzetta, Cooper} for
a clear presentation). 
Indeed, the description of a system in which
only the short wavelength  fluctuations are in thermal equilibrium can be
achieved by modifying the density matrix with respect to the thermal one, and
the CTP was designed just to describe systems with a generic density matrix. 

Anyway, the reader not familiar with the CTP formalism should not worry too
much about it. As we will discuss in Appendix A, the modification of the 
density
matrix that we will consider is equivalent to working in the more 
familiar RT formulation of
Niemi and Semenoff \cite{Niemi,Landsman} with a modified distribution 
function, given by
the Bose-Einstein distribution function (or the Fermi-Dirac, for
fermions) multiplied by the cutoff function. 

As a first stage of this program, in this paper we illustrate our method by 
considering the well-studied self-interacting real scalar theory.
As is well known, this model belongs to the universality class of the
Ising model, and then it has a second order phase transition. 
However, perturbation theory fails to reproduce this result even after the
resummation of daisy and super-daisy diagrams \cite{Espinosa}, unless the 
gap equations  are solved to $O(\lambda^2)$, $\lambda$ being the
quartic coupling constant \cite{Buchmuller}. 

We derive the ``exact'' Wilson RG equation for the effective potential 
and approximate it by expanding in powers of space-time  
derivatives of the field. As we will
discuss, even after our approximations, the effective potential
computed in this approach includes more contributions than the super-daisy
resummed one in \cite{Espinosa} and the prediction of
a second order phase transition will emerge in a clear way.
In particular, we will see that the running of the coupling constant, which is 
neglected in perturbation theory,  has a very strong effect. 
At the critical temperature, the effective coupling constant vanishes together
with the thermal mass, and this cures the severe infra-red problems encountered
in perturbation theory. 

The paper is organized as follows. 
In Section 2 we define the cutoff thermal field theory and discuss its
relationship with other RG approaches.
In Section 3 we derive the RG flow equations for 
the quantities of interest.
In Section 4 the evolution equation for the cutoff effective potential
is approximated in various ways in order to perform the numerical 
analysis.
Section 5 contains the results of our numerical study and Section 6 
some conclusions and further comments.
In Appendix A we briefly describe the Closed Time Path method and
how it applies to our case. Appendix B contains a discussion of the
problem of pinch singularities appearing in the kernel of RG flow
equations. 

\section{Cutoff thermal field theory}
As explained in the introduction, our aim is to define a 
coarse-graining procedure in which all the quantum fluctuations, 
and the high frequency thermal ones down to
an infra-red cutoff $\Lambda$, are integrated out.
In this section we will construct a path integral
representation of the resulting generating functional of 
coarse-grained Green functions,
in the case of a self-interacting real scalar field.

\subsection{Cutoff propagators}
Let us consider a free field, expanded in the usual manner 
in terms of annihilation and creation operators
\beq
\hat{\phi}(x) = \int \frac{d^3 k}{(2 \pi)^3 \:2 \omega_k} 
\left[ a_k \exp(-\i k\cdot x) + a_k^\dagger \exp(\i k\cdot x)\right],
\label{free}
\eeq
with $k_0=\omega_k=\sqrt{{\vec k}^2 + m^2}$; $m^2$ represents here
the zero-temperature, physical mass of the scalar particle.

In the case of an ideal gas in thermal equilibrium at a temperature 
$T=1/\beta$ one has \cite{Matsubara}
\beqra
\nonumber
\langle a_k^\dagger a_{k^\prime} \rangle_\beta  & = & (2 \pi )^3 2 
\omega_k N(\omega_k) \delta({\vec k} - {\vec k^\prime}), \\
&\label{equilibrium}\\
\langle a_k a_{k^\prime}^\dagger \rangle_\beta  & = & (2 \pi )^3 2 
\omega_k [1 + N(\omega_k)] \delta({\vec k} - {\vec k^\prime}), 
\nonumber
\eeqra
where $N(k_0)=[\exp(\beta k_0)-1]^{-1}$ is the Bose-Einstein 
distribution function.
The thermal averages in eqs.~\re{equilibrium} are both temperature-dependent. 
In particular, in the second line, we have a
``thermal'' contribution, the $N(\omega_k)$ inside the square
brackets, and a $T=0$, ``quantum'' contribution, the ``1''.
Of course, the commutator $[a_k, a_{k^\prime}^\dagger]$ being a
{\it c}-number, its expectation value is independent on the
state and is the same as in the zero-temperature vacuum
\beq
\langle [a_k, a_{k^\prime}^\dagger] \rangle_\beta =  (2 \pi )^3 2
\omega_k \: \delta({\vec k} - {\vec k^\prime}), 
\label{commutation}
\eeq
independent of $T$.
We propose to modify the above relations by introducing an infrared
cutoff {\it on the thermal part only}, that is
\beq
\begin{array}{ccl}
\langle a_k^\dagger a_{k^\prime} \rangle_\beta^\Lambda  & = &  
(2 \pi )^3 2 \omega_k N(\omega_k)  \Theta(|{\vec k}|,
\Lambda) \delta({\vec k} - {\vec k^\prime}), \\
&\\
\langle a_k a_{k^\prime}^\dagger \rangle_\beta^\Lambda  & = & 
(2 \pi )^3 2 \omega_k [1 + N(\omega_k)  \Theta(|{\vec k}|,
\Lambda) ] \delta({\vec k} - {\vec k^\prime}), 
\end{array}
\label{cutoff} 
\eeq
where, in general, $ \Theta(|{\vec k}|,\L)$ is a cutoff function 
which is $1$ for $|{\vec k}|\ge \L$ and rapidly vanishes for 
$|{\vec k}|<\L$.
The simplest choice for $\Theta(|{\vec k}|,\L)$ is
the Heavyside theta function
\beq\label{theta}
\Theta(|{\vec k}|,\L) \rightarrow \theta(|{\vec k}| - \L)\,.
\eeq
For simplicity, in the following we will always use the step 
function, even though in order to perform safe formal
manipulations one should choose \cite{Polchinski} a cutoff function which 
is always non-vanishing and of class $C^\infty$, 
and take only at the end the limit \re{theta}.

The physical state described by the relations in (\ref{cutoff}) is 
the one in which thermal equilibrium is achieved only by the fast, high
frequency, modes,  from $|{\vec k}|\rightarrow \infty$ down to the 
cutoff scale $|{\vec k}| \simeq \Lambda$.  The corresponding density matrix is
given by 
\beq
\rho = c \exp\left[-\int {\d}^3k \; \beta_{k,\L}\, a_k^\dagger a_k\right],
\label{densitym}
\eeq
where  $\beta_{k,\L}=\beta \omega_k$ for  
$|{\vec k}| > \Lambda$ and goes to infinity for  $|{\vec k}| < \Lambda$.
The quantum relation
\beq
\langle [a_k, a_{k^\prime}^\dagger] \rangle_\beta^\Lambda =  
(2 \pi )^3 2 \omega_k \: \delta({\vec k} - {\vec k^\prime}),
\eeq
analogous to eq.~\re{commutation}, is valid, in this state, for 
any value of the momenta, irrespective of the cutoff. Note that this 
would have not been the case if we had multiplied the whole right-hand
side of the second eq. \re{equilibrium} for the 
cutoff function $\Theta(|{\vec k}|,\L)$.

Now the cutoff Green functions are defined in the usual way, as
statistical averages of Heisenberg fields ordered along a  path $C$ in
the complex time plane \cite{Calzetta, Cooper}, running from $-\infty$ to $+\infty$
along the real time axis and back from $+\infty-i\varepsilon$ to
$-\infty-i\varepsilon$ infinitesimally below it.
As we will show in Appendix A, the effect of the non-thermal 
density matrix (\ref{densitym}) can be accounted for by simply working 
in the usual equilibrium
formalism, but with the 
Bose-Einstein distribution function  multiplied by the cutoff function.
For instance, the two-point Green function is
\beq
G^{(c)}(x, x^\prime)
= \theta_c(t-t^\prime) 
C^>(x, x^\prime)  +  \theta_c(t^\prime-t) C^<(x, x^\prime),
\label{green}
\eeq
where $\theta_c(t-t^\prime) $ is the generalization of the theta 
function on the contour $C$ and the two-point correlation functions 
are
\beq
C^>(x, x^\prime) = \langle \hat{\phi}(x) \hat{\phi}(x^\prime)
\rangle_\beta^\L = C^<(x^\prime,x)\,.
\label{corr}
\eeq
The free cutoff propagator 
$D_\L^{(c)}(x-x^\prime) =-\i G_0^{(c)}(x, x^\prime)$ 
can be computed in the standard way, 
by substituting (\ref{free}) in (\ref{green}) and \re{corr}, 
and using \re{cutoff}. One obtains
\beqra
&& \i D^{(c)}_\L (x-x^\prime) = \int \frac{{\d}^4 k}{(2 \pi)^4} 
2 \pi \delta(k^2 - m^2)
\e^{-\i k \cdot (x-x^\prime)} \nonumber\\
&& \quad\quad\times\left[\theta(k_0) \theta_c(t-t^\prime) + 
\theta(- k_0) \theta_c(t^\prime - t) +  N(|k_0|, \Lambda)\right],
\label{dc}
\eeqra
where
\beq\label{cutoffBE}
N(|k_0|, \Lambda) = N(|k_0|)\; \theta(|{\vec k}| - \Lambda)\,.
\eeq
The propagator can  be seen as a $2 \times 2$ matrix whose 
components are given by 
\beq
\begin{array}{cl}
D^{(11)}_\L(t-t^\prime) =& D^{(c)}_\L(t-t^\prime),\\
D^{(22)}_\L(t-t^\prime) =& D^{(c)}_\L((t-i\varepsilon)-(t^\prime - i\varepsilon)),\\
D^{(12)}_\L(t-t^\prime) =& D^{(c)}_\L(t-(t-i\varepsilon)),\\
D^{(21)}_\L(t-t^\prime) =& D^{(c)}_\L((t-i\varepsilon)-t^\prime),
\end{array}
\eeq
where $t$, $t^\prime$ lie on the real time axis and we have omitted 
the spatial arguments of the propagator.

Using expression \re{dc} and then Fourier transforming, we obtain 
the RT cutoff propagator in momentum space
\beq
D_\L(k) = \ds\left(
   \begin{array}{cc}
   \D_0 &  (\D_0 - \D_0^*)  \theta(-k_0) \\
   &\\
   (\D_0 - \D_0^*)  \theta(k_0)& - \D_0^*
   \end{array}
\right) 
+
(\D_0 - \D_0^*) \, N(|k_0|, \L)
\; B,
\label{propagator}
\eeq
where 
\beq
\D_0=\frac{1}{k^2 - m^2 + \i \varepsilon}
\eeq
and
\beq\label{B}
B=
\left(
   \begin{array}{cc}
   1  & 1  \\
   &\\
   1 & 1
   \end{array}
\right) \,.
\eeq
Note the separation between the $T=0$ and the thermal part.
In the $\varepsilon\rightarrow 0$ limit, we have
\beq
\D_0 - \D_0^* \longrightarrow -2\i \pi \delta(k^2-m^2),
\eeq
that is, the thermal part of the tree level propagator involves 
on-shell degrees of freedom only 
(only real particles belong to the thermal bath).

Since it will be useful in the following, we give here also the 
expression for the derivative of the propagator with respect to
the cutoff
\beq
\frac{\p\:\:}{\p \L} 
D_\L(k) = 2\i\pi \delta(k^2-m^2) \,
\delta(|{\vec k}| - \L) \, N(|k_0|)  \; B
\label{treederiv}
\eeq
and for the inverse of the propagator
\beqra
D_\L(k)^{-1} &=&  
\frac{1}{\D_0 \D_0^*} \left(
    \begin{array}{cc}
    \D_0^* & (\D_0-\D_0^*)
    \theta(-k_0) 
    \\&\\
    (\D_0-\D_0^*) 
    \theta(k_0) & -\D_0
    \end{array}
\right)\nonumber
\\
&&\nonumber\\
&-&\frac{\D_0-\D_0^*}{\D_0 \D_0^*}
N(|k_0|, \L)\left(
    \begin{array}{cc}
    1 & -1
    \\&\\
    -1 & 1
    \end{array}
\right)
\,.
\label{invprop}
\eeqra

\subsection{Cutoff effective action}
By introducing independent sources $j_1$ and $j_2$ for the two pieces of the
contour $C$ (the real time axis and the $t-i\varepsilon$ one, respectively),
the path integral representation
of the generating functional of RT cutoff Green functions 
can be written as \cite{Landsman}
\beq 
Z_\L[j] = \int [\d\phi_1] [\d\phi_2] \exp \i \left\{ \half 
\tr\, \phi \cdot D_\L^{-1} \cdot \phi + 
S[\phi] + \tr\, j \cdot \phi \right\},
\label{path}
\eeq
where the trace means integration over momenta while the dot
represents the sum over field indices
\beqra
\tr\,\phi \cdot D_\L^{-1} \cdot \phi &=& 
\sum_{a,b=1,2} \int \frac{{\d}^4 p}{(2 \pi)^4} \phi_a(p)\, 
[D^{-1}(p)]_{ab} \phi_b(-p), \nonumber\\
&& \label{dot} \\
\tr\, j \cdot \phi &=& \nonumber
\sum_{a=1,2}  \int \frac{{\d}^4 p}{(2\pi)^4} \, j_a(-p) \phi_a(p)
\eeqra
and $S[\phi] $ is the bare interaction action
\beq
S[\phi]=S[\phi_1]-S[\phi_2].
\eeq
Notice that in eq.~\re{path} it is assumed that the usual procedure of
renormalization of the perturbative ultraviolet divergences of the
zero-temperature theory has been carried out.
Namely we assumed a regulator, whose precise form is irrelevant for
our discussion, and a set of zero-temperature renormalization conditions.

In the usual interpretation of the RT formalism, $\phi_1$ and 
$\phi_2$ are respectively the ``physical'' field and the ``ghost'' 
field. 

{F}rom the expression of the propagator in (\ref{propagator}) one 
can immediately realize that, in the limit $\L\rightarrow\infty$
(which in practice, due to the exponential behaviour of the 
Bose-Einstein distribution function, means $\L\gg T$), none of the
thermal modes are propagating inside the loops. In this case by 
taking the functional derivatives of (\ref{path}) with respect to 
the source $j_1$ one obtains exactly the zero-temperature, fully 
renormalized, quantum field theory.
On the other hand, in the opposite limit $\Lambda \rightarrow 0$ the 
propagator tends to the equilibrium, Real Time, finite temperature
propagator, and consequently the generating functional in \re{path} 
gives the full finite temperature theory in thermal equilibrium. 

We define as usual the cutoff effective action as the Legendre 
transform of the generating functional of the connected Green 
functions $W_\L[j]= -\i \log Z_\L[j]$ \cite{Wetterich,Bonini,Morris}:
\beq
\half \tr\, \phi \cdot D_\L^{-1} \cdot\phi
+\G_\L[\phi] = W_\L[j] - \tr\, j\cdot\phi,
\quad\quad\quad\phi =\frac{\de W_\L[j]}{\de j}\,,
\label{action}
\eeq
where we have isolated the free part of the cutoff effective action 
and used for the classical fields the same notation as for the quantum 
fields.

As discussed in \cite{Niemi}, the free energy of the system is given by the
functional $\bar{\G}_\L[\varphi]$, defined by the relation
\beq
\frac{\delta \bar{\G}_\L[\varphi]}{\delta \varphi} =
\left.\frac{\delta \G_\L[\phi]}{\delta \phi_1}
\right|_{\phi_1=\phi_2=\varphi} \,.
\label{tadd}
\eeq
The tadpole $\bar{\G}_\L^{(1)}(\varphi)$, 
which will play an important r\^ole in the following, is found by
evaluating \re{tadd} for a space-time independent field configuration
\beq
\bar{\G}_\L^{(1)}(\varphi)=
\left.\frac{\delta \bar{\G}_\L[\varphi]}{\delta \varphi} 
\right|_{\varphi={\rm const.}}\,.
\label{tadd2}
\eeq
We conclude this subsection by recalling an important property.
By inspection, one finds that the cutoff effective action has a
discrete $Z_2$ symmetry \cite{Niemi}
\beq\label{Z2}
\G_\L[\phi_1,\phi_2]=-\G^*_\L[\phi_2^*,\phi_1^*]\,.
\eeq
This relation has the following consequences on the derivatives of the tadpole
\beq\label{Z2a}
\frac{\p \bar{\G}_\L^{(1)}(\varphi)}{\p\varphi}=
\left.\frac{\delta^2 \G_\L[\phi]}{\delta\phi_1^2}
\right|_{\phi_1=\phi_2=\varphi},
\eeq
\beq\label{Z2b}
\frac{\p^2 \bar{\G}_\L^{(1)}(\varphi)}{\p\varphi^2} 
= \sum_{a,b=1,2} \left.
\frac{\delta^3 \G_\L[\phi]}{\delta\phi_1 \delta\phi_a \delta 
\phi_b}\right|_{\phi_1=\phi_2=\varphi}
\,,
\eeq
where again $\varphi$ is space-time independent.
We will use these relations below.

\subsection{Relation to the other formulations of the Wilson RG}
Before proceeding with the discussion, we would like to comment on
the relationship between our approach and the other formulations of 
the Wilson RG existing in the literature.

The first application of the continuum Wilson RG philosophy to quantum 
field theory at $T=0$ is due to Polchinski \cite{Polchinski}. 
The motivation of that work was to provide a demonstration of perturbative 
renormalizability without resorting to diagrammatic techniques. 

In this formulation, which we will indicate in the following as 
the Polchinski RG, only the modes between an infra-red cutoff $\L$ and an
ultraviolet cutoff $\L_0$ are allowed to propagate inside loops.
The bare Euclidean Lagrangian depends only on the UV scale $\L_0$,
which is eventually sent to infinity, and the renormalization group 
flow describes the effect of quantum fluctuations 
coming into play as the infra-red cutoff $\L$ is lowered. 
This can be achieved in a way formally analogous to what we have
done in this section, the main difference being  in the form of the cutoff 
propagator. 
Namely, in this case the cutoff procedure has to be applied to the full 
propagator, and not only to a part of it, as we have done here. 
That is
\beq
D_{\L\L_0}(k) = \Theta(k,\L) \Theta(\L_0,k) D(k)\;,
\label{cutpol}
\eeq
where $D(k)$ is the tree level propagator. The internal legs of the Green 
functions generated by the analogues of our $Z_\L$, $W_\L$, and $\G_\L$, now
carry only momenta $\L<k<\L_0$. Therefore the 1PI Green functions are given
by the bare couplings for $\L=\L_0$ (since no modes are propagating
inside loops) and by the fully renormalized ones for $\L=0$ (since in this 
limit the quantum corrections have been included at any momentum scale). 

On the other hand, since the cutoff procedure described in this 
section modifies the thermal part of the propagator only, the Green 
functions defined in this paper contain the full quantum corrections,
independently of the value of $\L$. 
Lowering $\L$, we are introducing new {\it thermal} modes only.

In short, the Polchinski formulation of the Wilson RG  interpolates between
the {\it bare} and the {\it physical} (renormalized) theory, whereas the 
formulation presented in this paper interpolates between the {\it physical} 
theory at $T=0$ and the {\it physical} theory at $T\neq 0$. 

Since the two RG's describe two different physical problems, also the 
boundary conditions will be different. 
As stated previously, at $\L=\L_0$ in the Polchinski RG we have the bare 
Lagrangian. 
Thus, one should impose two different classes of boundary conditions
\cite{Polchinski,Bonini}. 
The first one involves all the ``irrelevant'' operators,
the ones with dimension larger than $4$. These should satisfy the
power counting in $\L_0$ at the ultraviolet scale $\L=\L_0$.
The remaining, ``relevant'', couplings have to be fixed by the matching 
of the renormalized theory and the physical observables, and so the initial
conditions for them should be given at $\L=0$.

On the other hand, in the approach presented in this paper, the physical
theory corresponds to the initial condition at $\L\to\infty$, so that all the 
couplings, both the relevant and the irrelevant ones, 
have to be fixed at this point. 

The Wilson RG method has already been applied to the 
study of the $T\neq 0$ quantum field theory in the IT formalism 
in ref.~\cite{Tetradis} and more recently in \cite{Liao}. 
The thermal propagator in the IT formalism \cite{Kapusta}
is obtained from the Euclidean propagator after the replacement 
$k_0\rightarrow\omega_n$, where $\omega_n=2 \pi T n$ are the Matsubara 
frequencies
\beq
\frac{1}{k^2+m^2} \rightarrow \frac{1}{(2 \pi T n)^2 + |{\vec k}|^2 + m^2}
\label{improp}
\eeq
and the $k_0$ integration is consistently replaced by $T \sum_n$. 

While in the RT formalism the quantum and the thermal part of the 
propagator can be clearly identified, as in (\ref{propagator}), in the 
Imaginary Time formalism this separation cannot be achieved, as we read 
from (\ref{improp}). 
So, in order to formulate a Wilson RG in this case, we can only modify 
the thermal propagator as in (\ref{cutpol}), imposing an IR cutoff on the 
combination $(2 \pi T n)^2 + |{\vec k}|^2$.

In this case we recover the bare theory for $\L =\L_0$ , since no
modes, either quantum or thermal, have been integrated in the loops, 
while for $\L=0$ we have the physical theory at $T\neq 0$. This formulation 
of the RG describes the effect of thermal and quantum fluctuations at the 
same time, and it is different from both  Polchinski's and ours. 
Concerning the boundary conditions in this case,  it should be noticed that 
since the RG now interpolates between the bare theory and the $T\neq 0$ 
physical theory, there is no value of $\L$ that corresponds to the physical 
theory at $T=0$, the one that is supposed to be directly related to the 
measurable observables. 
Then, in order to match the running parameters with the physical theory, 
a preliminary step has to be performed, consisting in the derivation of the 
bare parameters from the renormalized, $T=0$ theory 
\cite{Tetradis,Liao}.

\section{Renormalization group flow equations}
In this section we study the $\L$-dependence of the cutoff effective action.
By taking the derivative with respect to $\L$ of \re{path} 
we obtain the evolution equation for $Z_\L[j]$
\beq
\L\frac{\p\:\:}{\p\L} Z_\L[j] 
=-\frac{\i}{2} \tr\,\frac{\de \:\:}{\de j} \cdot 
\L \frac{\p\:\:}{\p \L} D_\L^{-1} \cdot 
\frac{\de \:\:}{\de j}  Z_\L[j] \,.
\label{evz}
\eeq
Choosing as initial conditions for $Z_\L[j]$ at 
$\L\gg T$ the full renormalized theory at zero 
temperature, the above evolution equation describes the effect of 
the inclusion of the thermal fluctuations at the momentum scale 
$|{\vec k}| = \L$. 

The evolution equation for the generating functional of the connected 
Green functions $W_\L[j]= -\i \log Z_\L[j]$ can
be derived straightforwardly from eq.~(\ref{evz})
\beq
\L\frac{\p\:\:}{\p\L} W_\L[j] = -\frac{\i}{2}  {\tr} \left[
\L\frac{\p\:\:}{\p\L} D_\L^{-1} \cdot 
\frac{\de^2  W_\L[j] }{\de j \:\de j} 
\right] 
+\frac{1}{2}{\tr}\,\frac{\de W_\L[j] }{\de j} 
\cdot \L\frac{\p\:\:}{\p\L} D_\L^{-1} \cdot
\frac{\de W_\L[j] }{\de j}\,. 
\label{evw}
\eeq
By using this equation in \re{action}, the flow equation
for the cutoff effective action $\G_\L[\phi]$ is found:
\beq\label{evact1}
\L\frac{\p\:\:}{\p\L} \G_\L[\phi] = \frac{\i}{2}  {\tr} \left[
\L\frac{\p\:\:}{\p\L} D_\L^{-1} \cdot 
\left(D_\L^{-1} +
\frac{\delta^2  \G_\L[\phi] }{\de\phi \:\de\phi} 
\right)^{-1} \right] \,.
\eeq
Notice that the terms coming from the $\L$-dependence of
the classical field $\phi$ cancel out, which is a well-known 
general property of the Legendre transform.
The evolution equations for the various vertices can be found by
expanding the r.h.s. in powers of $\phi$ \cite{Bonini,Morris}.

Deriving eq.~(\ref{evact1}) with respect to $\phi_1$ 
and then setting $\phi_1$ and $\phi_2$ equal to a constant background
$\varphi$, we
obtain the evolution equation for the tadpole, defined in eq.~(\ref{tadd2}),
\beq
\Lambda\frac{\partial\:\:}{\partial \Lambda} 
\bar{\Gamma}_\Lambda^{(1)}(\varphi) = 
-\frac{\i}{2} {\tr} \left\{
\left[ D_\Lambda^{-1}+ {\Sigma}_\Lambda
(\varphi)\right]^{-1} \cdot
\Lambda\:
\frac{\partial\:\:}{\partial \Lambda} {D}_\Lambda^{-1} 
\cdot 
\left[ {D}_\Lambda^{-1}+ {\Sigma}_\Lambda( 
 \varphi)\right]^{-1} \cdot
\Gamma_\Lambda^{(3)}(\varphi)
\right\}, 
\label{rtad}
\eeq
where 
\beq
\left[\G_\L^{(3)}(\varphi)\right]_{ij} =  
\left.\frac{\de\G_\L[\phi]}{\de\phi_i\:\de\phi_j\:\de\phi_1}
\right|_{\phi_1=\phi_2=\varphi}
\label{tri}
\eeq
and we introduced the field-dependent self-energy matrix
\beq
(2 \pi)^4 \delta^{(4)} (p + p^\prime) 
\left[{\Sigma}_\Lambda(p;\:\varphi)\right]_{ij} = \left.
\frac{\de^2\:\G_\L[\phi]}{\de\phi_i(p)\:\de\phi_j(p^\prime)}
\right|_{\phi_1=\phi_2=\varphi} \,.
\label{2free}
\eeq
First of all, we need to study the kernel of the evolution equation
for the tadpole, that is
\beq
{K}_\L(k;\:\varphi) \equiv 
- \i  \left[ {D}_\L^{-1}+ {\Sigma}_\L( 
\varphi)\right]^{-1} \cdot
\L\frac{\partial\:\:}{\partial \Lambda} {D}_\Lambda^{-1} 
\cdot 
\left[ {D}_\L^{-1}+ {\Sigma}_\L( 
\varphi)\right]^{-1}\,. \label{primeder}
\eeq
The kernel contains the ``full'' cutoff matrix propagator
$\left[{D}_\L^{-1}+{\Sigma}_\L(\varphi)\right]^{-1}$.
This can be obtained (by assuming a Schwinger-Dyson equation 
\cite{Landsman}) from eq.~\re{propagator} by substituting $\D_0$ with
\beq\label{DeltaL}
\D_\L = \D_0 \sum_{n=0}^{\infty}\left(- \Pi_\L \D_0\right)^{n} 
= \frac{1}{k^2 - m^2 + \Pi_\L(k;\:\varphi)
+ \i \varepsilon}
\eeq
where the quantity $\Pi_\Lambda(k;\:\varphi)$ is related to
the 11 component of the self-energy matrix by
\beq
\left\{
\begin{array}{ccl}
\ds {\Re}\,\Pi_\L(k;\:\varphi) &=&\ds  
{\Re}\,\left[{\Sigma}_\L(k;\:\varphi)\right]_{11}\\
&&\\
\ds {\Im}\,\Pi_\L(k;\:\varphi) &=& 
\ds \frac{1}{1 + 2 N (|k_0|, \L)} \,
{\Im}\,\left[{\Sigma}_\L(k;\:\varphi)\right]_{11}
\end{array}
\right.
\eeq
Collecting the above formulae we can compute the kernel, 
which turns out to be
\beq
{K}_\L(k;\:\varphi) =- \i (\D_0 - \D_0^*)
\frac{\D_\L \D_\L^*}{\D_0 \D_0^*} \: 
\L \,\de(|{\vec k}| - \L) \,N(|k_0|)
\; B
\,.
\label{kern}
\eeq
Notice that if one uses a cutoff function different from the step 
function, the kernel is obtained from \re{kern} by substituting 
$\de(|{\vec k}| - \L)$ with the $\L$-derivative of the new cutoff
function.

Comparing \re{kern} with the derivative with respect to $\Lambda$ of the 
tree level propagator, which we have computed in 
(\ref{treederiv}), we see that the delta function which forces the 
momenta on the zero temperature mass-shell has been replaced by
\beq
\i (\D_0 - \D_0^*)
\frac{\D_\L \D_\L^*}{\D_0 \D_0^*}
=- 2 \pi  \rho_\L(k; \varphi) +\i \D_0 \D_0^* (\Pi_\L - \Pi_\L^*) 
(1 + R_\L + R^*_\L
+R_\L R^*_\L) , 
\label{nk}
\eeq
where 
\beq
\rho_\L(k; \varphi) = -\frac{\i}{2 \pi} \left[\D_\L - \D_\L^*\right]
\label{spectral}
\eeq
is the interacting spectral function and
\[
R_\L = \sum_{n=1}^{\infty}\left(\Pi_\L \D_0\right)^n.
\]
The second contribution to the kernel in eq.~(\ref{nk}) exhibits pinch
singularities of the form $\D_0^m \; {\D_0^*}^n$. It is also
proportional to the imaginary part of the self-energy $\Pi_\L(k)$, which
is non-vanishing along the whole real axis in the $k_0$ complex
plane\footnote{We thank S. Jeon for
drawing our attention on this point.}. As will be shown in Appendix B,
these pinch singularities cancel with analogous contributions coming
from $\Gamma^{(3)}_\L$ once the kernel (\ref{nk}) is inserted in the
evolution equation for the tadpole to give
\beq
\L\frac{\p\:\:}{\p\L} 
\ds \bar{\G}^{(1)}_\L(\varphi)=-\i \frac{\L}{2}
\int\frac{{\d}^4 k}{(2\pi)^4}\:\de(|{\vec k}| - \L)
(\D_0-\D_0^*) \frac{\D_\L \D_\L^*}{\D_0 \D_0^*}
\,N(|k_0|) \bar{\G}^{(3)}_\L(k; -k; \varphi)\; ,  
\label{evpot1}
\eeq
where 
\beq
\bar{\G}^{(3)}_\L(k; k'; \varphi) \equiv \frac{\p^2
\bar{\G}^{(1)}_\L(\varphi)}{\p\varphi(k) \p \varphi(k')} 
= {\Tr} \left\{{B}\cdot
\G^{(3)}_\L(k;k'; \varphi)\right\}
\label{mir}
\eeq
(here the trace is over $1,2$ field indices)
and $\G^{(3)}_\L$ has been defined in (\ref{tri}). 

It is important to stress here that the above equation is exact, since
no approximation such as perturbative expansion or truncation has been
performed up to now.

In the rest of this section we will discuss the evolution equation
obtained neglecting the imaginary part of the self energy on-shell, which in
perturbation theory arises only  at two-loops \cite{Parawani, Jeon1}.
We will use this approximation as the starting point for the
numerical analysis which  we will describe in sect. 4.
In this approximation, the kernel in (\ref{nk}) can be written as
\beq
\frac{\i}{k^2 - m^2 + {\Re}\,\Pi_\Lambda(k;\:\varphi) + 
\i \varepsilon} - 
\frac{\i}{k^2 - m^2 + {\Re}\,\Pi_\Lambda(k;\:\varphi) - 
\i \varepsilon},
\label{diffe}
\eeq
so that the only singularities are the poles in the $k_0$ complex 
plane satisfying $k_0^2 = |{\vec k}|^2 + m^2 
-{\Re}\,\Pi_\Lambda(k;\:\varphi) \pm \i \varepsilon =0$, 
and it vanishes in the rest of the complex plane. Notice that the
pinch singularities have disappeared in this approximation.

In the case in which
$|{\vec k}|^2 + m^2 -{\Re}\,\Pi_\L(k;\:\varphi)>0$ 
the poles lie infinitesimally close to the real 
axis, and (\ref{diffe}) reduces to $ 2 \pi \delta( k^2 - m^2 + 
{\Re}\,\Pi_\Lambda(k;\:\varphi))$. 
On the other hand, if  $ |{\vec k}|^2 + m^2 -{\Re}\,\Pi_\L
(k;\:\varphi) < 0$, as  can be the 
case when there is spontaneous symmetry breaking, the poles lie on 
the imaginary axis, and 
the contributions of the two pieces in (\ref{diffe}) to the
integration along the real $k_0$ axis cancel. 

Then, the quantity (\ref{nk}) appearing in the kernel reduces, 
in the $\varepsilon\rightarrow 0$ limit, to
\beq
\i (\D_0 - \D_0^*)
\frac{\D_\L \D_\L^*}{\D_0 \D_0^*}
 \rightarrow 2 \pi \delta( k^2 - m^2 + {\Re}\,\Pi_\Lambda(k;\:\varphi))
\,
\theta(|{\vec k}|^2 + m^2 -{\Re}\,\Pi_\Lambda(k;\:\varphi))
\,.
\label{simpli}
\eeq
The physical meaning of this kernel is straightforward. As the 
infrared cutoff $\Lambda$ is lowered, the mass associated to
the new modes coming into thermal equilibrium is not given by the 
$T=0$ pole mass $m^2$, but by the thermal mass, that is
the pole of the full propagator obtained by integrating over the high 
momentum modes ($|{\vec k}|>\Lambda$), already in thermal
equilibrium. This comes out quite naturally in this formalism, 
whereas in perturbation theory one has to perform ad hoc
resummations in order to cure the infrared divergences.
 We will come back to the comparison 
between this approach and perturbation theory in the
following. 

The other interesting feature of this kernel has to 
do with the case of spontaneous symmetry breaking. 
In perturbation theory, one is not allowed to compute 
the effective potential for values of the background field
close to the symmetric phase, if the temperature is less than the 
critical temperature of the phase transition. This is because
the thermal mass squared is negative, in this region, which gives 
rise to a complex effective potential. As showed by  Weinberg and Wu  
\cite{Weinberg}, the real part of this effective potential can still 
be interpreted
as the energy density of a spatially homogeneous, although unstable, state. 

On the other hand, in
this approach, we see that the only modes giving rise to a thermal 
evolution are those with a real energy, and the running
stops as soon as the energy squared  becomes negative, so that no imaginary 
parts for the effective potential are generated.

Using \re{simpli}, the evolution
equation 
for the tadpole (\ref{evpot1}) now takes the
simple form, 
\beqra
&&\L\frac{\p\:\:}{\p\L} 
\ds \bar{\G}^{(1)}_\L(\varphi)=-\pi\L
\int\frac{{\d}^4 k}{(2\pi)^4}\:\de(|{\vec k}| - \L)
\de(k^2-m^2+{\Re}\,\Pi_\L(k;\:\varphi^\prime))\,N(|k_0|) \nonumber
\\ &&\qquad\times
\theta(|{\vec k}|^2 + m^2 -{\Re}\,\Pi_\L(k;\:\varphi)) 
\;\bar{\G}^{(3)}_\L(k; -k; \varphi)\; ,
\label{evpot2}
\eeqra
which will be studied numerically in the next section in an
approximation scheme based on a derivative expansion and truncations.

\begin{figure}[ht]
\begin{center}
\mbox{\epsfig{file=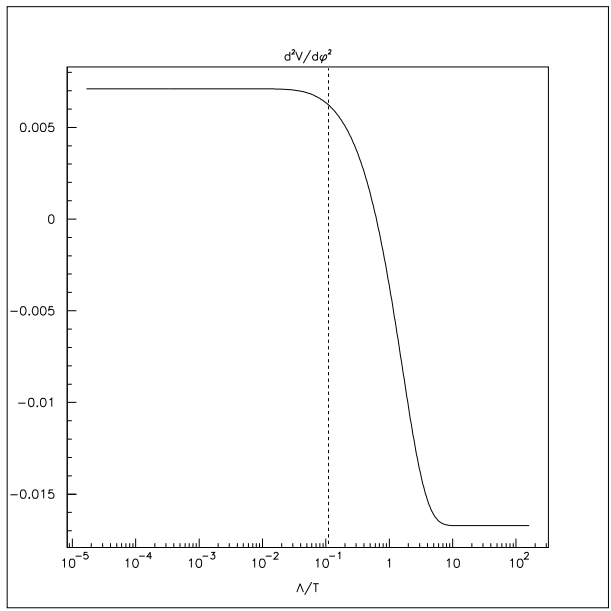,height=8cm}}
\end{center}
\caption{The running of the thermal mass
($V''_\Lambda(\varphi=0)$, here given in units of $v^2$, $v$ being the zero
temperature vev) with respect to the ratio $\Lambda/T$. The region to 
the left of
the dashed line corresponds to the three-dimensional regime (see text). We
have fixed  $T=1.2 \:T_c$ and $\lambda= 0.1$.}
\label{1}
\end{figure}

Before concluding this section, we would like to briefly discuss the infra-red 
limit of the evolution equation for the tadpole (\ref{evpot2}).
As the energy of the thermal modes becomes much smaller than $T$,
the Bose-Einstein distribution function can be approximated as
\beq 
N(|k_0|) \simeq \frac{T}{|k_0|} 
\left(1 - \frac{1}{2} \frac{|k_0|}{T} + \ldots \right).
\label{approx}
\eeq
Integrating in  $k_0$ in (\ref{evpot1}) and keeping only the first 
term in the expansion for $N(|k_0|)$ in  (\ref{approx}) the evolution 
equation for the tadpole becomes
\beq
\L\frac{\p\:\:}{\p\L} 
\bar{\G}^{(1)}_\L(\varphi)\simeq -
\frac{\L\: T}{2}
\int\frac{{\d}^3 k}{(2\pi)^3}\:
\de(|{\vec k}| - \L)
\frac{\bar{\G}^{(3)}_\L(\vec{k}; -\vec{k}; \varphi)}
{|{\vec k}|^2+m^2+{\Re}\,\Pi_\L(\vec{k};\:\varphi)}
\label{dimrid}
\eeq
if 
\beq
\omega_\L^2 = \L^2 + {\Re}\,
\Pi_\L(k_0=\omega_\L,|\vec k |=\L;\:\varphi) \ll T^2.
\label{limit}
\eeq
Equation~(\ref{dimrid}) shows that, in the limit of eq.~(\ref{limit}),
and neglecting the imaginary part of the self-energy, the exact 
four-dimensional running at finite temperature reduces to a purely 
three-dimensional 
one, at zero temperature. The momentum integral is now a genuine
three-dimensional loop integral with the zero-temperature propagator.
Equation ~(\ref{dimrid}) is, in fact, the same equation as would have been 
obtained in the Polchinski RG in three dimensions and $T=0$,
provided the following matching between the three-dimensional and
the four-dimensional couplings had been done:
\beqra
\bar{\G}^{(1)}_\L(\varphi)_{D=3} &=&
\frac{1}{\sqrt{T}}\,\bar{\G}^{(1)}_\L(\varphi)_{D=4}\nonumber
\\&&\nonumber\\
{\Re}\,\Pi_\L({\vec k};\:\varphi)_{D=3}&=&
{\Re}\,\Pi_\L(k_0=\omega_k,{\vec k};\:\varphi)_{D=4}\\
&&\nonumber\\
\bar{\G}^{(3)}_\L({\vec k};\;-{\vec k}; \varphi)_{D=3}&=&
\sqrt{T}\,\bar{\G}^{(3)}_\L(k_0=\omega_k,{\vec k};\;
k_0^\prime=\omega_k,-{\vec k};\;\varphi)_{D=4} \nonumber \,.
\eeqra
Obviously, the above matching has to be performed at a value of $\L$ such 
that the condition (\ref{limit}) is fulfilled, that is in the 
three-dimensional regime. 
So, the question arises whether we can neglect, in a first approximation, 
the four-dimensional running from $\L=\L_0 \gg T$ to some value of $\L$, 
$\L_{3D}$, which is inside the three-dimensional regime. 
The answer is clearly negative, as we can read from Fig.~\ref{1}.

In this picture, we have plotted the value of the thermal mass in 
$ \varphi=0$ (see Sect.~5) as a function of the ratio $\L/T$. 
The area to the left of the dashed line represents the region in which the 
Bose-Einstein 
distribution function can be approximated by $T/|k_0|$ with an accuracy 
better than $10 \%$, and consequently the
three-dimensional RG equation (\ref{dimrid}) is valid to that accuracy.

As we can see, most of the running of the mass, from the zero-temperature
value (corresponding to $\L/T\gg 1$) to the finite temperature one, takes 
place before the three-dimensional region is reached. 
In other words, in order to match the three-dimensional couplings with the 
physical four-dimensional ones, it is essential to keep the pure 
four-dimensional
running (at $T\neq 0$) under control, and the study of the pure 
three-dimensional running is, in general, not justified.

On the other hand, dimensional reduction is justified when 
one is interested in the study of the critical theory. 
By universality, we know that this theory is insensitive to the ``initial 
conditions'' of the RG equations, so that the matching with the physical 
theory at zero temperature is not very important for the computation
of universal quantities.

In our numerical analysis we will 
always consider the four-dimensional RG equations derived in the 
previous sections, without dimensional reduction. 
This approach will allow us  to study both  the universal quantities
(critical exponents) and the non-universal ones (critical temperature, 
thermal masses), and to keep the transition from the 
four-dimensional to the effectively three-dimensional critical theory 
under control.
At the end of sect.~5 we will comment on the possibility of using a 
dimensionally-reduced form of eq.~\re{evpot1}  in order to study
critical regime of the theory.

\section{Derivative expansion, truncations, and comparison with perturbation
theory}
An analytic solution of the exact evolution equation for the tadpole in 
eq.~(\ref{evpot1}) is not available. 
In principle one has to know the momentum
dependence of both the full self-energy and the vertex
$\bar{\G}_\L^{(3)}$,  i.e. of the first and second derivative of the 
tadpole with respect to the field $\varphi$. 
A similar problem is encountered  in the applications of the
Polchinski RG in the zero-temperature field theory. 
An approximation scheme based on a
derivative expansion (or momentum-scale expansion \cite{derexp}) has 
proved to be very efficient in that context, at least in the case of the scalar
theory.

In order to perform a systematic derivative expansion in this case, 
we expand the free energy functional defined in (\ref{tadd}) 
in derivatives of the field $\varphi(x)$ as follows:
\beq
\bar{\G}_\L[\varphi] = \int {\d}^4x\left[ \half m^2 \varphi^2 - V_\L(
\varphi) + \half (\partial \varphi)^2 Z_\L(\varphi) +
\half (u \cdot \partial \varphi)^2 Y_\L(\varphi) + \cdots\right]
\label{derexp}
\eeq
where the dots indicate higher derivative terms. Note the term containing the
quadrivector of the thermal bath, $u_\mu$, which has to be introduced for
relativistic covariance. 

Stopping the expansion in (\ref{derexp}) at some order, plugging it into eq.
(\ref{evpot1}), and equating the coefficients of the terms with 
the same powers of derivatives of the field, we obtain evolution
equations for the functions $V_\L$, $Z_\L$, $Y_\L$, and higher orders.

A remark is in order. Usually in the Polchinski RG 
the expansion \re{derexp} is
not well defined if one uses a sharp cutoff function. The problem arises
because in the zero external momentum limit the analogue of our 
eq.~(\ref{evact1}) and its momentum derivatives contain ill-defined 
products of $\delta(|{\vec k}|- \L)$, coming from the derivative of
the cut-off propagator, and $\theta(|{\vec k}|- \L)$, coming from the 
underived propagator \cite{derexp}.
Fortunately, this is not the case in our approach. The point is,
again, that the theta
function here appears only in  the thermal part of the propagator, and all the
above mentioned products of theta and delta functions cancel one another by
the same mechanism that ensures the cancellation of the ``pinch'' singularities
 in the RT
formalism. As a consequence, the sharp cutoff limit can safely be taken even at
higher orders in the derivative expansion.

At the lowest order, the derivative expansion corresponds to 
neglecting the momentum
dependence of the self-energy and the three-point vertex appearing in
the r.h.s. of \re{evpot1}.
In this approximation we can use \re{Z2a} in \re{DeltaL} and $\D_\L$ 
can be written as $\D_\L=[k^2-V''_\L(\varphi)]^{-1}$, where the 
prime over the effective potential indicates derivation with respect 
to the field. Analogously the three-point vertex in \re{mir} is equal
to $-V'''_\L(\varphi)$. Note that since the imaginary part of the
self-energy vanishes for zero external momenta, it does not contribute
to the zero order in the derivative expansion. We get
\beq
\L\frac{\p\:\:}{\p\L} 
V^\prime_\L(\varphi)=-\frac{\L^3}{4\pi^2}\:
\frac{N(\omega_\L(\varphi))}
{\omega_\L(\varphi)}\,V'''_\L(\varphi)\,
\theta(\Lambda^2 + V_\L''(\varphi))
\,,
\label{pde}
\eeq
where 
\[
\omega_\L(\varphi) = \sqrt{\L^2+V''_\L(\varphi)}\; .
\]
Using the equilibrium Bose-Einstein distribution for $N$
and integrating in $\varphi$ we find the following evolution 
equation for the effective potential
\beq
\Lambda \frac{\partial\:\:\:}{\partial \Lambda}
V_\Lambda(\varphi) = 
- T \frac{\Lambda^3}{2 \pi^2} \log\left[1-\exp\left(-\beta 
\sqrt{\Lambda^2 + V_\Lambda''(\varphi)}
\right)\right]\,
\theta(\Lambda^2 + V_\L''(\varphi)) .
\label{heat}
\eeq
The same equation was found in the Matsubara formalism in \cite{Liao2}.

Looking at the effective potential as a function of the two variables 
$\Lambda$ and $\varphi$ we see that the
evolution equation is a non-linear partial derivative differential 
equation, with the initial condition
$V_{\Lambda=\infty}(\varphi)$ being the renormalized effective
potential at zero temperature.

In principle, one could seek for a numerical solution of the 
evolution equation (\ref{pde}). An alternative procedure, which will 
be followed in the present paper, is to take further derivatives of 
eq.~(\ref{pde}) with respect to the field $\varphi$, so that the 
partial derivative differential equation is turned into a infinite 
system of ordinary first order differential equations in $\L$ for the 
unknowns $V_\L'$, $V_\L''$, $V_\L''',$ $\ldots$, with
$\varphi$-dependent coefficients,
\beq
\left\{
\begin{array}{lcl} \ds
\L\frac{\p\:\:}{\p\L} 
V^\prime_\L(\varphi)&=&\ds -\frac{\L^3}{4\pi^2}\:
\frac{N(\omega_\L)}
{\omega_\L}
\,V'''_\L(\varphi)\,
\theta(\Lambda^2 + V_\L''(\varphi))\\
&&\\
\ds \L\frac{\p\:\:}{\p\L} 
V''_\L(\varphi)&=& \ds-\frac{\L^3}{4\pi^2}
 \:
\left\{
\frac{N(\omega_\L)}
{\omega_\L}
\,V''''_\L(\varphi)\,
+\frac{d\;}{d V''_\L} \left[\frac{N(\omega_\L)}
{\omega_\L}\right]
\left[V'''_\L (\varphi)\right] ^2 \right\}
\:
\theta(\Lambda^2 + V_\L''(\varphi)) \\
&&\\
&\equiv& -\L 
{\cal F}_\L \left[ V_\L''(\varphi)\right] 
\\
&\cdot &\\
&\cdot &\\
&\cdot &\\
\end{array}
\right.
\label{syst}
\eeq
In the following we will solve this system of equations at different 
orders of truncations, in order to test the reliability of this 
further approximation (besides the derivative expansion). 

This also allows us to make a comparison between the present approach 
and the perturbative one. The 1-loop perturbative result \cite{Dolan}
and the daisy and super-daisy \cite{Espinosa,Buchmuller,Sdaisy} 
resummed ones correspond to different approximations of the above system,
 truncated at most to second order, 
 i.e. without taking into account
the evolution of the third and higher order couplings. More precisely:

{\bf (i) 1-loop perturbation theory} corresponds to a truncation to 
the first equation only,  the one for the tadpole, in which $V_\L''$ and 
$V_\L'''$ are approximated by their tree level values:
\beq
V_\L''(\varphi) \simeq m^2(\varphi),\:\:\:\:\:
V_\L'''(\varphi) \simeq \lambda \varphi \:\:;
\eeq

{\bf (ii) Daisy resummation}  improves the 1-loop result by
partially taking 
into account the evolution of the ``mass'' $V_\L''$, replacing
$m^2(\varphi)$ with a  $\L$-independent ``thermal mass'' in the evolution
equation for the tadpole. The trilinear
and quadrilinear couplings are approximated by the tree level ones, that is 
\beq
\begin{array}{c}
\ds V_\L''(\varphi) \simeq m^2_T(\varphi)=m^2(\varphi) +
\int_0^\infty d\L'{\cal F}_{\L'} \left[m^2(\varphi) \right]\\
\\
\ds V_\L'''(\varphi) \simeq \lambda \varphi \;\;\;\;\;\;\;\;\;\;\;\;\;\;
V_\L''''(\varphi) \simeq \lambda 
\end{array}
\eeq 
where the function ${\cal F}_\L $, defined in (\ref{syst}), is computed
by approximating $V_\L''(\varphi) $ with the tree level mass.

{\bf (iii) Super-daisy resummation}  corresponds to an improvement of
the previous approximation consisting in using as the value of the thermal mass
the solution $\bar{m}_T$ of the ``gap equation''\footnote{Different resummation
schemes have been presented in the literature, which however agree with 
one another
up to $O(\beta)$, where $\beta=\lambda T/m_T$ is the effective expansion
parameter.}
\beq
\begin{array}{c}
\ds V_\L''(\varphi) \simeq \bar{m}^2_T(\varphi)=
m^2(\varphi) +
\int_0^\infty d\L'{\cal F}_{\L'} \left[ \bar{m}^2_T(\varphi) \right]\\
\\
V_\L'''(\varphi) \simeq \lambda \varphi\;\;\;\;\;\;\;\;\;\;\;\;\;\;
V_\L''''(\varphi) \simeq \lambda .
\end{array}
\eeq
As we see, the (resummed) perturbative results do not take into account the
evolution equations for the trilinear and higher orders couplings. 
Moreover, the
$\L$ dependence of the second derivative of the effective potential is not
considered, while a $\L$-independent thermal mass is introduced. 

In the next section, we will see how these effects are in general very 
important, mainly at
temperatures close to the critical one. In particular, the inclusion 
of the running
up to the fourth derivative of the potential, will change the transition
from a (weakly) first-order one to a second-order one.

\section{Numerical results}
We have solved the system in eq.~(\ref{syst}) at different orders of 
truncation
and for different values of $\varphi$. In this way we are able to reconstruct
the shape of the tadpole, and then of the effective potential. 
Since we are mainly
interested in the study of the phase transition, we will solve the evolution
equations for $\varphi$ close to the origin, and running from 
$\L=\L_0 \gg T$
(in practice $\L_0 > 10 T$ will be enough, see Fig.~\ref{1})
down to $\L=0$. As initial condition in $\L=\L_0$ we use the 1-loop effective
potential at $T=0$,
\beq
V_{\L_0}(\varphi) = -\frac{1}{2} \mu^2 \varphi^2 +\frac{\lambda}{4 !} 
\varphi^4 +
\frac{m^4(\varphi)}{64 \pi^2}\left[ \log \frac{m^2(\varphi)}{-2\mu^2} - 
\frac{3}{2}\right]\;,
\eeq
where
\[m^2(\varphi) = -\mu^2+\frac{\lambda}{2} \varphi^2\;,
\]
and the potential has been renormalized in the $\overline{MS}$ scheme with
$Q^2=-2\mu^2$. In principle, different approximation schemes to the $T=0$
effective potential could also be used, as given for instance by the 
Polchinski RG
or by lattice computations. However, due to universality, the results for the
critical theory should be almost insensitive to the details of the 
initial conditions,
and the 1-loop approximation will be enough for our present purposes.

\begin{figure}[bh!]
\begin{center}
\mbox{\epsfig{file=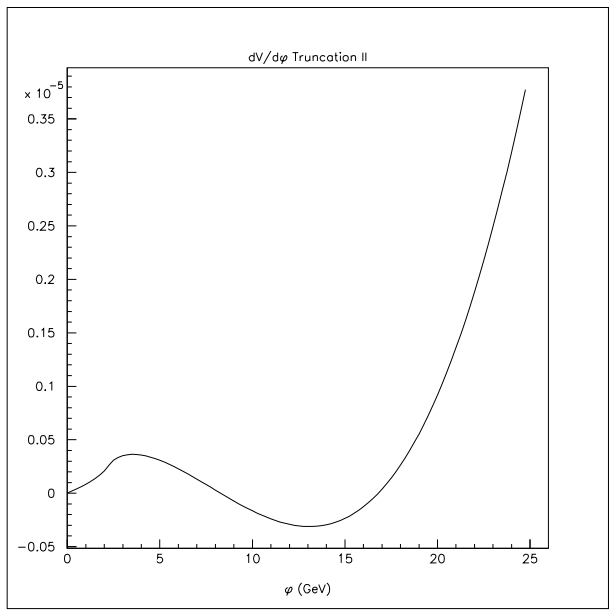,height=8cm}}
\end{center}
\caption{The tadpole $V'_{\Lambda=0}$ (in units of $v^3$) as a function
of $\varphi$, obtained by truncating to order II. The temperature
has been chosen such that the effective potential has two degenerate minima.
We have fixed $v=246$ GeV and $\lambda=0.1$.}
\label{2}
\end{figure}

\begin{figure}[t]
\begin{center}
\mbox{\epsfig{file=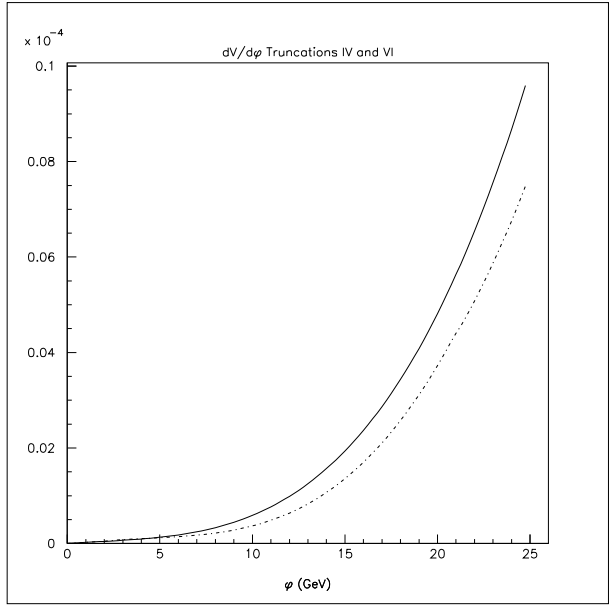,height=8cm}}
\end{center}
\caption{The tadpole $V'_{\Lambda=0}$ (in units of $v^3$) as a function
of $\varphi$, obtained by truncating to order IV (continuous line) and VI
(dash-dotted line).
The temperature corresponds to the vanishing of the second derivative
at the origin. We have fixed $v=246$ GeV and $\lambda=0.1$.}
\label{3}
\end{figure}

\begin{figure}
\begin{center}
\mbox{\epsfig{file=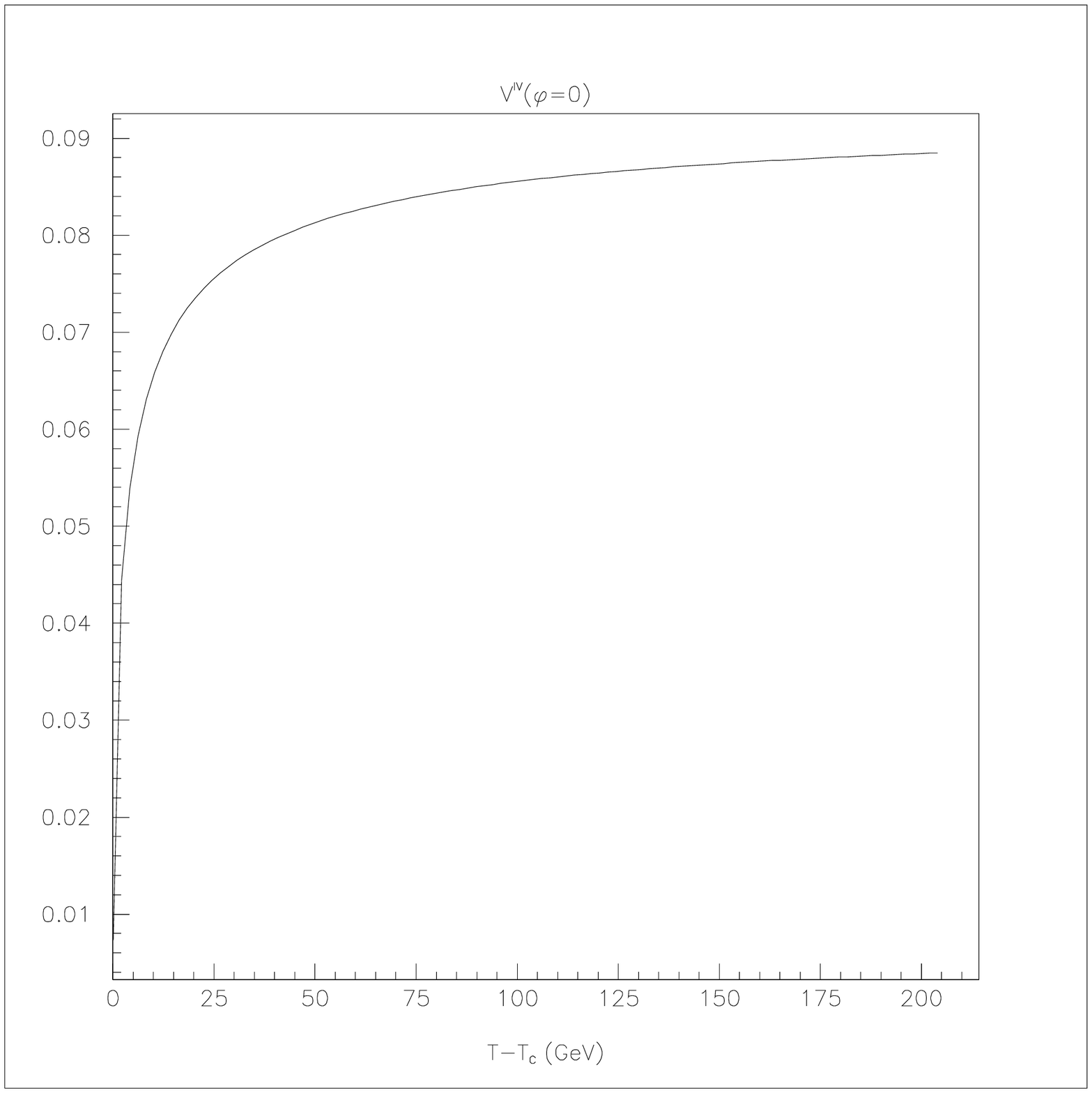,height=8cm}}
\end{center}
\caption{The running of the coupling constant
$V''''_{\Lambda=0}(\varphi=0)$ as a function of the temperature, 
for $T \ge T_c$.
We have fixed $v=246$ GeV and $\lambda=0.1$.}
\label{4}
\end{figure}

In Fig.~\ref{2} we show the tadpole $V'_{\L=0}$ as a function of $\varphi$ 
at the critical temperature, when the system has been truncated to the second
line. As we see the potential has three
stationary points (the zeros of $V'_{\L=0}$),  i.e. there are two
minima, so that it describes a (weakly) first-order phase transition.
In this case, we have
fixed the temperature in such a way that the two minima are degenerate. This
result agrees with the one obtained in 1-loop resummed perturbation
theory, see for instance ref.~\cite{Espinosa}. 

However as soon as we turn  the evolution for
the third and fourth coupling on, as we did in Fig.~\ref{3},
we see that things change dramatically. 
In this case the tadpole has only one zero in $\varphi=0$ for temperatures
higher than a critical value and two zeros for lower temperatures.
There is no temperature interval in which a third zero is present,
in other words the effective potential develops no barrier between the
symmetric and the asymmetric phases.
The critical temperature $T_c$  has been
determined by requiring that the second derivative of the potential 
vanished. 
So, as soon as the running of the coupling is taken into account, the 
transition turns
out to be of  second-order, as would be expected from universality. The
continuous and dash-dotted lines in this plot represent the results obtained
by truncating to the fourth and sixth orders, respectively.

The running of
the coupling is indeed a strong effect, in particular close to the critical
temperature, as we read from Fig.~\ref{4}. 
The $T=0$ value for the coupling in this
picture is chosen to be $0.1$, and we see that, while for $T$ far from 
$T_c$ the
running is an $\sim 10 \%$ effect, for $T\rightarrow T_c$ the coupling
vanishes. 

The expansion parameter of the super-daisy resummation is
given by $\beta=\lambda T/m_T$, where $m_T$ is the mass scale of 
the theory, in
this case the thermal mass $V''_{\L=0}$. As $T\rightarrow T_c$ the 
thermal mass
vanishes, and $\beta$ diverges. This is the source of the wild infra-red 
problems of
resummed perturbation theory, which prevent computations for temperatures
close to the critical one. Including the running of the coupling, which 
is also
vanishing at $T_c$, $\beta$ tends to a finite value, and the infrared 
problems are, if
not cured, at least domesticated. This observation was already
made in ref. \cite{Tetradis}. 

Looking at the evolution equations, we see that the range of $\L$ for 
which there
is a sizeable running of the parameters is determined by the function
$N(\omega_\L)/\omega_\L$. For $\L \gg T$ the distribution function is
exponentially suppressed and the running is negligible,
as we read from Fig.~\ref{1}. High energy modes are Boltzmann suppressed. 

On the other hand, for small $\L$, the function goes approximately as 
$T/(\L^2 + V''_\L)$, and so there is running as long as $\L^2 > V''_\L$. 
We can
then define a correlation length as the inverse of $\L_c$, such that
\beq
\L_c^2=V''_{\L_c}\:.
\eeq
Since there are no thermal modes at
wavelengths larger than the correlation length, integrating over 
$\L < \L_c$ gives a negligible effect.

At the critical temperature we have $V''_{\L=0}=0$, and the
correlation length diverges. 
We can define critical indices in the usual way 
(see for instance ref.~\cite{Parisi}) and see how their computation is
improved with respect to perturbation theory in the present approach.

At the lowest order in the derivative expansion, we have computed 
the critical exponent $\nu$, governing the scaling of the
renormalized mass near the critical temperature,
\beq 
m_{\L=0} \sim |T-T_c|^\nu\; ,
\eeq 
and $\delta$, describing  how the magnetization at the
critical temperature, $M(h, T=T_c)=\overline{\phi}$, 
scales with the magnetic field $h=V'_\L$,
\beq
M(h, T=T_c)=\overline{\phi}\sim |h|^{1/\delta} = (V'_\L)^{1/\delta}.
\eeq
\begin{table}
\begin{center}
\begin{tabular}{||c|c|c|c|c|c|c||}\hline
{\rm Truncation} & $\lambda$ & $T_c/v$ & {\rm Order trans.} & $\nu$ & $\delta$ 
&$\eta$ \\
\hline\hline
{\rm II}        & 0.1         &  1.99   &        {\rm I}          &   -- &    --  & --\\
\hline
{\rm IV}      & 0.1         &   2.01   &        {\rm II}        &   0.53& 3.27 &-- \\
\hline
{\rm IV+Z$_{\rm II}$}& 0.1     &  2.02     &       {\rm II}       & 0.53 &  3.27  & 0.015\\
\hline
{\rm VI}        & 0.1      &   2.02     &      {\rm II}       & 0.58  & 3.57 & --\\
\hline
Best values \cite{Parisi,Zinn}&           &              &       {\rm II}       
& 0.63  & 4.82 & 0.032\\
\hline
\end{tabular}
\caption{Results for the order of 
the phase transition, the critical temperature,
and the critical exponents $\nu$, $\delta$, and $\eta$ at different orders of
approximations to the evolution equations (see text). The last row contains the
best values in the literature.}
\label{tab1}
\end{center}
\end{table}
\newline
In Table ~\ref{tab1} 
we summarize our results for the order of the phase transition, the
critical temperature, and the critical exponents $\nu$ and $\delta$ 
obtained at the lowest order in the derivative expansion, and
truncating the system in eq.
(\ref{syst}) at second, fourth, and sixth order respectively. 

In the last row we have listed  the best results for the critical 
exponents obtained in the literature \cite{Parisi,Zinn}. 
We see that passing from
the fourth to the sixth order of truncation we have a sizeable 
improvement in the result for $\nu$ and $\delta$.

Including the corrections of $ O\left((\partial \varphi)^2\right)$,  
keeping also the third and fourth terms in the expansion in (\ref{derexp}),
we can compute the critical exponent $\eta$, which describes the scaling of the
two-point renormalized Green function with respect to $\L$ for 
$\L \rightarrow 0$ at $T=T_c$:
\beq 
\overline{\Gamma}^{(2)}_\L(0) = m^2_\L (1+Z_\L) \sim \L^{2-\eta}\;.
\eeq
Taking again derivatives with respect to the field $\varphi$, we
obtain a system of
first order differential equations for $V'_\L$, $V''_\L$, $\ldots$, 
$Z_\L$, $Z'_\L$, $\ldots$, $Y_\L$, $Y'_\L$, $\ldots$. 
We have truncated this system at the fourth
derivative for $V_\L$ and at the second one for $Z_\L$ and $Y_\L$. Due to the
$Z_2$ symmetry of the model, and using the fact that the initial condition for
$Y_\L$ (which corresponds to its zero-temperature value) is zero, the effect of
$Y_\L$ and its derivatives is negligible for values of $\varphi$ 
close to the origin, and can be neglected with respect to that of $Z_\L$. 
Moreover, we have still neglected the imaginary part of the
self-energy on-shell, which is only a two-loop effect in standard
perturbation theory \cite{Parawani, Jeon1}.

The result is reported in the fourth row of Table ~\ref{tab1}. As we
can read, the effect of wave function renormalization is very small, as may be
expected since in perturbation theory it arises at two loops. 
So, in this model, the
derivative expansion turns out to be a very efficient approximation scheme. 

The accuracy of our results for the critical exponents is generally 
a lot better than the one obtained in other approaches to the 
four-dimensional theory, such as perturbation theory or the IT 
formulation of the RG \cite{Tetradis}, both predicting $\nu=0.5$. 

Nevertheless, the best way to compute the critical exponents
remains the study of the three-dimensional critical
theory (see refs.~\cite{Parisi,Zinn} and, for a RG computation,  
\cite{TetradisII} ). 
Once one has established that the effective potential solution of eq.~\re{heat}
describes a second order phase transition, the universal quantities
(the critical exponents) can be obtained studying the evolution around
the fixed points of the critical ($T=T_c$) theory at large length-scales 
($\L\to 0$). 
For $\L\ll T_c$ one expects that the theory becomes scale invariant.
This is what indeed happens, as can be seen by rescaling 
$V_\L$ and $\varphi$ according to $V_\L=\L^3 T_c \,\tilde V_\L$ and 
$\varphi=\sqrt{\L T_c} \,\tilde\varphi$
(notice that the temperature enters as expected by dimensional
reduction arguments). In terms of the new variables, eq. \re{heat} becomes
\[
\frac{\partial}{\partial t}\tilde V_\L +\half
\tilde\varphi \tilde V'_\L -3 \tilde V_\L=
\frac{1}{2\pi^2} \log\left[1-\exp\left(-{\e}^{-t}\sqrt{1+\tilde
V''_\L}\right)\right]\,,\qquad t=\log\left(\frac{T_c}{\L}\right)\,.
\]
As $t\to \infty$ the explicit $t$-dependence in the  RHS
disappears, and we obtain  the scale-invariant equation
(dropping tildes and changing variables to
absorb the factor in front of the logarithm)
\beq\label{LPA}
\frac{\partial}{\partial t} V_\L +\half
\varphi V'_\L -3 V_\L=\log\left(1+V''_\L\right)\,.
\eeq
This is the RG flow equation for the zero temperature potential in $3$
space-time dimensions at the lowest order in the derivative expansion 
and for a sharp momentum cutoff \cite{Morris3}. We see than that the critical
behaviour of \re{heat} is effectively described by a three dimensional
$T=0$ theory. The fixed points and the corresponding critical indices
of eq.~\re{LPA} can be computed numerically without truncations and
the result for $\nu$ is (see \cite{Morris3} for details)
\[
\nu=0.6895\,.
\]
On the other hand, if one is interested in studying the theory out of the 
critical regime, or in the case of a first order phase transition, 
the three-dimensional approach is no longer suited, and the full 
four-dimensional theory has to be addressed. In this case, as we 
have discussed at the end of sect.~3, the present formulation of the
Real Time RG provides a clear connection between the theory at finite
temperature and the theory at $T=0$, which  we are supposed to test in the 
laboratory.

\section{Conclusions and outlook}
In thermal field theory  several difficulties are encountered while 
performing perturbative calculations. 
These problems come from the severe infra-red divergences 
which plague finite temperature Green's functions, in particular in 
gauge theories. To this aim a technique for the resummation 
of the so-called ``hard thermal loops'' was proposed and
developed \cite{Kobes}. 
However, this method is not successful for scales in which 
non-perturbative physics (such as the gluon magnetic mass in QCD) 
become relevant. 
The key problem is that ordinary perturbation theory does not clearly 
separate the various scales in the game, and the diagrammatic expansion
has to be reorganized.

We believe that
the Wilson RG approach should be very helpful in this sense.

In this paper we have considered the Wilson RG formulation of
a RT thermal field theory. 
The main idea is to consider the thermal interaction between a 
quantum field and a thermal bath as an effective interaction, 
namely to regard the frequency modes above a certain scale $\L$ as 
effective interactions for the low energy modes below $\L$. 
Therefore the thermal fluctuations above $\L$ are integrated out 
and put in a ``Wilsonian effective action'',
which satisfies an ``exact'' (in principle) evolution equation in 
$\L$.
We solved this flow equation for the effective potential 
in various approximations, providing a non-perturbative resummation 
of Feynman graphs.
The approach is physically quite transparent, rigorous, and gives 
better numerical 
results with respect to usual resummed perturbation theory. 
The Wilson method had already been applied to
the same self-interacting scalar model in the Imaginary Time Matsubara 
formalism of thermal field theories \cite{Tetradis,Liao}.
Our results agree with these previous analyses and are better 
in some cases. 
However, the main aim of this paper was to set up the general
formalism and discuss some approximation methods, by using the 
well-studied scalar theory. 

Concerning the extension of this RT approach to gauge theories,
we believe it is very promising, since the issue of gauge invariance
can be considered in a clean way by using the CTP formalism described
in the Appendix A. 
Moreover, if one is interested in the static
quantities (such as effective potentials and thermal masses), the
imaginary parts of the self-energies can be neglected and only
on-shell modes contribute to the flow, as shown by \re{simpli}. In
this case,
by imposing the thermal boundary conditions only on the physical 
degrees of freedom, as suggested in \cite{Landshoff}, one finds that the 
thermal on-shell part of the propagators is gauge-invariant. Therefore 
the RG evolution equations obtained in this way provide a gauge-invariant 
resummation. The consequences of this observation will be considered 
in a separate paper.

The RT formalism is absolutely necessary in order to 
deal with non-equilibrium phenomena. 
We hope this paper is a first step towards a consistent 
formulation of coarse-graining in non-equilibrium field theory.
As far as the problem of thermalization is concerned, it should 
be possible to derive a Boltzmann equation, at least for a 
quasi-equilibrium or quasi-stationary system, and give a precise
physical meaning to the IR cutoff $\L$, which could be interpreted as
the thermalization scale, $\L(t)$, to be determined dynamically.
With such a systematic 
treatment, one should be able to compute corrections to the Boltzmann 
equation, at least in some regimes.
More generally, the dynamical elimination of the degrees of freedom 
which are irrelevant at a certain scale, would permit to derive an 
effective theory (for instance a kinetic theory \cite{Jeon}) from the 
underlying fundamental field theory.
These points are currently being explored.

\noindent
{\bf Acknowledgements}

We are very grateful to M. Bonini and T.R. Morris for helpful advices
and reading of the manuscript. We also thank T.R. Morris for having
suggested us the dimensional reduction argument reported at the end of
sect.~5.
We thank for discussions
W. Buchm\"uller, D. Comelli, S. Jeon, H.J. de Vega and G. Marchesini.
M. D'A. would like to thank the LPTHE (Paris VI-VII) for hospitality and 
the foundation Della Riccia for partial support.
M.P. acknowledges support from the EC ``Human Capital and Mobility'' 
programme.

\appendix

\section{The Closed Time Path formalism}
In this Appendix, we give a short review of the CTP formalism \cite{noneq}, 
and discuss its application to the formulation of the Wilson 
RG presented in this paper. We will
follow the presentation of refs. \cite{Calzetta} and \cite{Cooper} (see
also the review \cite{Boyanovsky}).

This technique was introduced to describe the 
evolution of an arbitrary initial state, described
by a density matrix $\rho$. 

In the usual path integral approach to
scattering theory, we are interested to the transition amplitudes
between a $|in\rangle$ and a $|out\rangle$ vacuum state in the presence
of an external source. We now face  a different physical problem.
We want to follow the out-of-equilibrium time evolution of an 
interacting field, without knowing its state in the far future.
Moreover, we look for a real and causal evolution equation. For a
Hermitian field operator, the diagonal matrix element
$\langle in| in\rangle$ is  real and the off-diagonal ones 
$\langle out|in\rangle$ are complex.

Therefore the basic idea of the CTP method is to start from a diagonal
$\langle in| in\rangle$ matrix element at a given time $t=0$ and 
insert a complete set of states at a later time $t'$:
\beq
Z[J_+,J_-]=\int [{\d}\psi] \langle in,0|\psi,t'\rangle_{J_-}
\langle\psi,t' | in,0\rangle_{J_+}\;,
\eeq
where $J_+$ and $J_-$ are the two sources for the propagation 
forward and backward in time, respectively.
Notice that one puts different sources for the two transition
amplitudes, since one is mainly interested in following the forward 
time evolution.
By introducing the density matrix $\rho$ of the initial state and 
setting $t'=+\infty$ one gets
\begin{eqnarray*}
&&Z[J_+,J_-,\rho]=\int [{\d}\varphi][{\d}\varphi'][{\d}\psi]
\langle \varphi,0 | T^* \exp \left(-\i\int J_- \Phi\right)
|\psi,+\infty\rangle
\\ && \quad\times
\langle \psi,+\infty | T \exp \left(\i\int J_+ \Phi\right)
|\varphi',0\rangle
\langle \varphi',0 | \rho|\varphi,0\rangle\;,
\end{eqnarray*}
where $T$ and
$T^*$ are the time and anti-time ordering operators, respectively.
We see the origin of the name CTP. This path-integral is
performed on a contour in the complex time plane running from $t=0$ to
$t=\infty$ and then back to $t=0$.
One can then
introduce a standard path integral representation for the two matrix
elements corresponding to the transitions from the state at $t=0$ to
the state at $t=t'$ and viceversa:
\beq
Z[J_+,J_-,\rho]=\int [{\d}\phi_+][{\d}\phi_-]
\langle \phi_+,0 | \rho|\phi_-,0\rangle
\exp\i\left\{S[\phi_+]+J_+\phi_+-S^*[\phi_-]-J_-\phi_-\right\}\,.
\eeq
It is clear how the $+,-$ fields and sources correspond to the ones
labelled with $1,2$ in the paper, respectively. Therefore we
introduce the vectors
\beq
\phi_a=(\phi_+,\phi_-)\;,
\qquad
J_a=(J_+,J_-)\;.
\eeq
As far as the initial state is concerned,
one can always decompose the density matrix element at $t=0$ in the
following way:
\beq\label{densmatrix}
\langle \phi_1,0 |\rho|\phi_2,0\rangle = \exp\i\left\{
K+\int K_a\phi_a+\half\int K_{ab}\phi_a\phi_b+\ldots\right\}
\,.
\eeq
Therefore the information on the initial state is all contained in the
non-local sources $K$, which are obviously concentrated at $t=0$.
The various $K$'s are just boundary conditions in time for the
corresponding Green functions. In this way a perturbative expansion
may be constructed, with the usual Feynman rules.

We now specify the form of the initial state. 
For an initial state of a real free particle in thermal equilibrium,
$\rho\sim\exp[-\beta\int \d^3 k\;\omega_k a^\dag_k a_k]$ and the matrix
element is \cite{Calzetta}:
\beq\label{dmeq}
\langle \phi_1,0 |\rho|\phi_2,0\rangle \sim \exp
-\half\int\frac{\d^3 k}{(2\pi)^3}\frac{\omega_k}{\sinh\beta\omega_k}
[(\phi_1^2(k)+\phi_2^2(k))\cosh(\beta\omega_k) - 2\phi_1(k)\phi_2(k)]
\;.
\eeq
We see that the effect of this initial state is just to add a
correction to the free propagators of the theory. In this way the free
propagator of the RT thermal field theory is obtained. 
For a interacting  particle, the matrix element
in \re{densmatrix} cannot be computed exactly, but the Wick theorem
may be used to perform the usual perturbative calculations of the RT 
equilibrium theory.
In this way we find the same result as we would have found by applying
the well-known property that, for a pure thermal
state, the matrix element $\langle \phi_1,0 |\rho|\phi_2,0\rangle$ 
admits a path integral representation over the path going from $t=0$ 
along the imaginary time axis to the time $t=-\i\beta$. 
It may also be shown that indeed there is a certain freedom in the 
choice of the path to go from $t=0$ to $t=-\i\beta$, so that all the
various formulations of the equilibrium thermal field theory are obtained.

The density matrix relevant to this paper, eq.~(\ref{densitym}), is obtained
by introducing a momentum-dependent temperature 
\beq
\beta_{k,\L}=\left\{
\begin{array}{cc}
\beta \omega_k & {\rm for}\;\; |\vec{k}|>\L\\
&\\
+\infty & {\rm for}\;\; |\vec{k}|<\L
\end{array}
\right. \;.
\eeq
As discussed in \cite{Calzetta}, if the  density matrix has this form 
all the initial $n$-point correlations,  
K in \re{densmatrix}, vanish for $n>2$, and  the result for
the density matrix element \re{densmatrix} is obtained from  
eq.~\re{dmeq} after the substitution $\beta\omega_k\to\beta_{k,\L}$. 

As a consequence, the only modification with respect to the usual 
equilibrium RT
formalism will consist in replacing the Bose-Einstein distribution function in
the free propagators, with the  cut off one defined in \re{cutoffBE}.

\section{Cancellation of pinch singularities}
In this appendix we discuss the mechanism of cancellation of the pinch
singularities which appear in the matrix kernel $K_\L$ 
\re{primeder}, which we rewrite in the following way
(we suppress the index $\L$ and restore the field indices 1,2 
for sake of notation)
\beq
\frac{\partial\:\:}{\partial\L} \G^{(1)}_1=\half \tr
K_{ab}\G^{(3)}_{1ab}\,.
\label{tadeq2}
\eeq
We will be able to rewrite this equation in a way which is manifestly
free of the pinch-like singularities $\D_0^m {\D_0^*}^n$ appearing in
eq. \re{nk}.

First of all recall that the full matrix propagator 
$G_\L=[D_\L^{-1}+\Sigma_\L]^{-1}$ is free of pinch singularities due to 
the matrix structure of $D_\L$ and $\Sigma_\L$ \cite{Landsman}, and so it
is its $\L$-derivative.
Then the origin of the pinch singularities in the kernel of the RG
flow equation \re{tadeq2} is clear if one realizes that the definition 
\re{primeder} corresponds to the $\L$-derivative of the full propagator
taking into account only the $\L$-dependence coming from the {\it free} 
propagator $D_\L$. 
This destroys the matrix structure of $G_\L$. In order to recover it 
(and therefore to cancel the pinch singularities), from the vertex
$\G^{(3)}_{1ab}$ in \re{tadeq2} there should come a contribution like 
\[
-G_\L \cdot \left(\frac{\partial\:\:}{\partial\L}\Sigma_\L \right) 
\cdot G_\L \,,
\]
which, added to the kernel \re{primeder}, would give the total (pinch-free) 
$\L$-derivative of $G_\L$. 

Indeed this happens in the following way. 
The $\L$-derivative of the matrix self-energy is found by derivating
with respect to $\phi_i$, $\phi_j$ the evolution equation \re{evact1}
\beq
\frac{\partial\:\:}{\partial\L}\Sigma_{ij}=\half \tr
K_{ab}[\G^{(4)}_{ijab}-2\G^{(3)}_{ica}G_{cd}\G^{(3)}_{jdb}]
\,.\label{seflow}
\eeq
Then one has the identity
\beq
\i\G^{(3)}_{1ij}=\i\tilde\G^{(3)}_{1,ij}+\tilde\G^{(3)}_{1,ab}
G_{aa'}G_{bb'}[\half\G^{(4)}_{a'b'ij}-\G^{(3)}_{a'ci}G_{cd}\G^{(3)}_{b'dj}]
\,,\label{tpr}
\eeq
where $\tilde\G^{(3)}_{k,ij}$ is a vertex which is 
{\em two-particle-irreducible} if one tries to separate the external 
fields $ij$ from the external field $k$. 
By inserting this equation in \re{tadeq2}
and using \re{seflow} one has
\[
\frac{\partial\:\:}{\partial\L} \G^{(1)}_1=\half\tr
\left( \i\frac{\partial\:\:}{\partial\L} G_{ab} \right)
\tilde\G^{(3)}_{1,ab}\,,
\]
which does not exhibit pinch singularities.
We can summarize this mechanism of cancellation as follows: in the
product $K_{ab}\G^{(3)}_{1ab}$ the
contributions corresponding to the $\L$-derivative of the matrix
self-energy come from the parts of $\G^{(3)}_{1ij}$ in which the
external fields
$ij$ are two-particle-reducible with respect to the external field $1$ (the
reader may convince himself of this by drawing the corresponding
diagrams). These two-particle-reducible parts are the last term in the
r.h.s. of eq.~\re{tpr}. Then the $\L$-derivative of the matrix self-energy
cooperates with the kernel $K_\L$ to give the $\L$-derivative of the
full propagator, which is free of pinch singularities.


\begin{thebibliography}{99}

\bibitem{Wilson} 
K.G. Wilson, {\em Phys. Rev. B\/} {\bf 4} (1971) 3174, 3148;
K.G. Wilson and J.G. Kogut, {\em Phys. Rep.\/} {\bf 12} (1974) 75.

\bibitem{Niemi} 
A. Niemi and G. Semenoff, {\em Ann. of Phys.\/} {\bf 152} (1984) 105;
for a review and a list of references see ref.~\cite{Landsman}.

\bibitem{Matsubara}
T. Matsubara, {\em Prog. Theor. Phys.\/} {\bf 14} (1955) 351.

\bibitem{noneq}
J. Schwinger, {\em J. Math. Phys.\/} {\bf 2} (1961) 407;
P.M. Bakshi and K.T. Mahanthappa, {\em J. Math. Phys.\/} {\bf 4}
(1963) 1 and {\bf 4} (1963) 12;
L.V. Keldysh, {\em JETP} {\bf 20} (1965) 1018;\\
K. Chou, Z. Su, B. Hao and L. Yu, {\em Phys. Rep.\/} {\bf 145} (1987) 
141.

\bibitem{Calzetta}
E. Calzetta and B. L. Hu, {\em Phys. Rev. D\/} {\bf 35} (1988) 495 and
{\bf 37} (1988) 2878.

\bibitem{Cooper}
F. Cooper, S. Habib, Y. Kluger, E. Mottola, J.P. Paz and R. Anderson,
{\em Phys. Rev. D\/} {\bf 50} (1994) 2848.

\bibitem{Landsman}
See for instance N.P. Landsman and Ch.G. van Weert, 
{\em Phys. Rep.\/} {\bf 145} (1987) 141.

\bibitem{Espinosa} 
J.R. Espinosa, M. Quiros and F. Zwirner, {\em Phys. Lett. B\/} 
{\bf 291} (1992) 115.

\bibitem{Buchmuller}
W. Buchm\"uller, Z. Fodor, T. Helbig and D. Walliser, {\em Ann. Phys.\/} 
{\bf 234} (1994) 260. 

\bibitem{Polchinski} 
J. Polchinski, {\em Nucl. Phys. B\/} {\bf 231} (1984) 269.

\bibitem{Wetterich} 
C. Wetterich, {\em Nucl. Phys. B\/} {\bf 352} (1991) 529.

\bibitem{Bonini} 
M. Bonini, M. D'Attanasio and G. Marchesini, {\em Nucl. Phys. B\/}
{\bf 409} (1993) 441.

\bibitem{Morris} 
T.R. Morris, {\em Int. J. Mod. Phys. A\/} {\bf 9} (1994) 2411.

\bibitem{Parawani}
R.R. Parawani, {\em Phys. Rev. D\/} {\bf 45} (1992) 4695.

\bibitem{Jeon1} 
S. Jeon, {\em Phys. Rev. D\/} {\bf 47} (1993) 4586.

\bibitem{Tetradis} 
N. Tetradis and C. Wetterich, {\em Nucl. Phys. B\/} {\bf 398} (1993) 
659.

\bibitem{Liao} 
S.-B. Liao and M. Strickland, {\em Phys. Rev. D\/} {\bf 52} (1995) 
3653.

\bibitem{Kapusta} 
J.I. Kapusta, {\em Finite temperature field theory\/}
(Cambridge University Press, Cambridge, 1989).

\bibitem{Weinberg} E.J. Weinberg and A. Wu, {\em Phys. Rev. D\/} 
{\bf 36} (1987) 2474.  

\bibitem{derexp} 
T.R. Morris, {\em Nucl. Phys. B\/} {\bf 458[FS]} (1996) 477
and references therein.

\bibitem{Liao2} 
S.-B. Liao, J. Polonyi and D.-P. Xu, {\em Phys. Rev. D\/} {\bf 51} 
(1995) 748. 

\bibitem{Dolan} 
L. Dolan and R. Jackiw, {\em Phys. Rev. D\/} {\bf 9} (1974) 3320.

\bibitem{Sdaisy} 
M.E. Carrington,  {\em Phys. Rev. D\/} {\bf 45} (1992) 2933; 
M. Dine, R.G. Leigh, P. Huet,\\
 A. Linde and D. Linde,  
{\em Phys. Rev. D\/} {\bf 46} (1992) 550; 
C.G. Boyd, D.E. Brahm and S.D. Hsu, {\em Phys. Rev. D\/} {\bf 48} (1993) 4963;
P. Arnold and O. Espinosa, {\em Phys. Rev. D\/} {\bf 47} (1993) 3546.

\bibitem{Parisi} 
G. Parisi, {\em Statistical field theory\/} (Addison-Wesley, Redwood 
City, 1988).

\bibitem{Zinn} 
J. Zinn-Justin, {\em Quantum field theory and critical phenomena\/} 
(Clarendon, Oxford, 1989).

\bibitem{TetradisII} 
N. Tetradis and C. Wetterich, {\em Nucl. Phys. B\/} {\bf 422[FS]} (1994) 
541.

\bibitem{Morris3}
T.R. Morris, {\em Phys. Lett. B\/} {\bf 334} (1994).

\bibitem{Kobes}
For a recent review and a list of references see 
R. Kobes, {\em Hard thermal loop resummation techniques in hot gauge
theories\/}, hep-ph/9511208.

\bibitem{Landshoff}
P.V. Landshoff and A. Rebhan, {\em Nucl. Phys. B\/} {\bf 383} (1992) 607
and  {\rm ERRATUM}{\em - ibid.\/} {\bf 406} (1993) 517.

\bibitem{Jeon}
S. Jeon and L.G. Yaffe, {\em From quantum field theory to 
hydrodynamics: Transport coefficients and effective kinetic theory\/},
hep-ph/9512263.

\bibitem{Boyanovsky}
D. Boyanovsky, H.J. de Vega and R. Holman, 
Nonequilibrium dynamics of phase transitions: from the early Universe
to chiral condensates,
in: H. J. de Vega and \\N. S\'anchez, eds., 
{\em Proceedings of the Second Paris Cosmology Colloquium\/}
(World Scientific, Singapore, 1995) 127--215.

\end{thebibliography}
\end{document}